\documentclass[a4paper,10pt]{article}
\usepackage[english]{babel}
\usepackage[T1]{fontenc}
\usepackage[utf8]{inputenc}
\usepackage{amssymb}
\usepackage{physics}
\usepackage{amsmath}
\usepackage{graphicx}
\usepackage[unicode]{hyperref}
\usepackage{tikz}
\usepackage{color}
\usepackage{genyoungtabtikz}
\usepackage[title]{appendix}
\usepackage{indentfirst}
\usepackage{bm}
\usepackage{xhfill}
\usepackage{bbm}
\usepackage{multirow}
\usepackage{array}
\usepackage[centertableaux]{ytableau}
\usepackage{multicol}
\usepackage[shortlabels]{enumitem}
\usepackage{forest}
\usepackage{float}
\usepackage{mathrsfs}

\usetikzlibrary{decorations.pathreplacing}

\usepackage{ytableau}

\hypersetup{
    colorlinks=true,
    linkcolor=blue,
    filecolor=magenta,      
    urlcolor=cyan,
}
 
\usepackage{amsthm}

\theoremstyle{definition}

\usepackage[nottoc]{tocbibind}

\usepackage{authblk}

\title{\textbf{Virasoro Zero-Mode Flow in Critical Topologically Massive Gravity: Logarithmic Correlators and Partition Function Grading}}

\author[1,2]{\textbf{Yannick Mvondo-She}}
\affil[1]{National Institute of Theoretical and Computational Sciences, Private Bag X1, Matieland, South Africa}
\affil[1]{\texttt{yannick.mvondo-she@nithecs.ac.za}}
\affil[2]{Department of Physics, University of Pretoria, Private Bag X20, Hatfield 0028, \hspace{1cm} South Africa}

\date{}

\begin{document}

\maketitle

\begin{abstract}
We study the rank-two logarithmic sector of critical Topologically
Massive Gravity through the Virasoro zero-mode action
\[
U(s,\bar s)=e^{-sL_0-\bar s\bar L_0}.
\]
Writing \(L_0=h\mathbf 1+\frac{1}{2}N\) and
\(\bar L_0=\bar h\mathbf 1+\frac{1}{2}N\), with \(N^2=0\), separates the
semisimple conformal grading from the nilpotent mixing of the
logarithmic pair. 

We apply the Virasoro zero-mode action to logarithmic two-point
functions and to the conformal grading of the logarithmic
contribution to the one-loop partition function. For the correlators,
the semisimple part generates the ordinary conformal power law, while
the nilpotent mixing governed by the combination \(s+\bar s\) produces
the characteristic logarithmic dependence. For the critical-TMG
weights \((h,\bar h)=(2,0)\), this gives a holomorphic power law
together with logarithmic dependence on the full Euclidean
separation. For the partition function, the relations \(q=e^{-s}\)
and \(\bar q=e^{-\bar s}\) express its conventional multiplicative
grading in terms of the additive parameters of the same Virasoro
zero-mode action, providing an exact reparameterization of the established one-loop result.

The two applications capture complementary information: logarithmic
correlators resolve the off-diagonal Jordan mixing, whereas the
ordinary graded trace organizes conformal weights, multiplicities,
and descendant content without separately resolving the nilpotent
matrix element. The Virasoro zero-mode action therefore
provides a unified representation-theoretic description of these
boundary structures in critical TMG.
\end{abstract}

\tableofcontents

\section{Introduction}
\label{sec:introduction}

Three-dimensional gravity acquires propagating local degrees of
freedom when the Einstein--Hilbert action with negative cosmological
constant is augmented by a gravitational Chern--Simons term
\cite{Deser:1981wh,Deser:1982vy}. The resulting theory,
Topologically Massive Gravity (TMG), provides a useful setting for
examining the interplay between bulk dynamics, asymptotic conformal
symmetry, and holography. For an \(\mathrm{AdS}_3\) background with
radius \(\ell\), the structure of the linearized spectrum is governed
by the dimensionless parameter \(\mu\ell\), where \(\mu\) is the
topological mass.

A new perspective on TMG emerged with the proposal of Ref.~\cite{Li:2008dq} that the theory acquires special properties at the chiral point
\begin{equation}
\mu\ell=1.
\end{equation}
At this point, the massive graviton degenerates with the left-moving
gravitational mode, and it was proposed that, subject to
Brown--Henneaux boundary conditions, the problematic bulk excitation
could be removed so as to leave a chiral theory with only the
right-moving boundary sector. This proposal initiated an extensive
investigation of the spectrum, stability, boundary conditions, and
holographic interpretation of critical TMG
\cite{Carlip:2008jk,Giribet:2008bw}.

A central development was the explicit construction of the
logarithmic mode by Grumiller and Johansson
\cite{Grumiller:2008qz}. At the degeneracy point, the ordinary
left-moving solution is accompanied by a generalized solution with
logarithmic asymptotic behaviour. The status of this mode depends
crucially on the choice of boundary conditions. Under strict
Brown--Henneaux asymptotically \(\mathrm{AdS}_3\) boundary conditions
\cite{Brown:1986nw}, the logarithmic mode is excluded, leading to the
chiral-gravity sector. The chiral theory and its gravitational path
integral have subsequently motivated broader developments in
three-dimensional quantum gravity
\cite{Maloney:2015ina,Eberhardt:2022wlc}. 

A different theory results when the asymptotic conditions are relaxed
so as to retain the logarithmic solution. In this setting, critical
TMG admits a logarithmic graviton whose asymptotic behaviour violates
the standard Brown--Henneaux falloff only logarithmically
\cite{Grumiller:2008qz}. The resulting theory, commonly referred to
as \emph{log gravity} \cite{Maloney:2009ck}, is the theory considered
in the present work. Its consistency and holographic structure have
been investigated extensively
\cite{Grumiller:2008pr,Grumiller:2008es,Grumiller:2009mw,
Ertl:2010dh,Gaberdiel:2010xv}, leading to the proposed interpretation
of critical TMG with logarithmic boundary conditions as a
gravitational dual of a logarithmic conformal field theory
\cite{Grumiller:2010rm,Grumiller:2010tj,Grumiller:2013at}.
Holographic renormalization at the chiral point reproduces the
characteristic logarithmic two-point functions and new anomaly of the
dual LCFT \cite{Skenderis:2009nt}, while complementary bulk
calculations yield the same logarithmic correlation structure
\cite{Grumiller:2009mw}.

Logarithmic conformal field theories arise when the action of the
conformal generators becomes non-diagonalizable, giving rise to
indecomposable representations and logarithmic correlation functions
\cite{Gurarie:1993xq,Flohr:2001zs}. Logarithmic structures were
considered in the context of AdS/CFT
\cite{Kogan:1999bn} and subsequently in
\(\mathrm{AdS}_3/\mathrm{LCFT}_2\) holography
\cite{Myung:1999nd}, providing a broader setting for the logarithmic
modes that appear at critical points of three-dimensional gravity.
At the representation-theoretic level, such structures are described
by indecomposable Virasoro modules, including staggered modules whose
extension structure, descendant spaces, and logarithmic couplings
have been studied systematically \cite{Kytola:2009ax}. The rank-two
Jordan pair arising in critical TMG may therefore be viewed as a
gravitational realization of this general LCFT representation
structure.

The logarithmic sector also has a nontrivial state-counting structure.
Following the one-loop analysis of log gravity
\cite{Gaberdiel:2010xv}, the logarithmic contribution to the
partition function was reformulated in terms of Bell polynomials in
Ref.~\cite{Mvondo-She:2018htn}, and subsequent work developed
connections with Schur-polynomial structures, Hilbert series,
symmetric products, integrable KP hierarchies, and Hurwitz-type
constructions
\cite{Mvondo-She:2019vbx,Mvondo-She:2021joh,Mvondo-She:2022jnf}.
These developments provide the state-counting background for the
partition-function discussion below.

For the logarithmic graviton of critical TMG, the left-moving mode
\(\psi_{\mu\nu}^{L}\) and its logarithmic partner
\(\psi_{\mu\nu}^{\log}\) form a rank-two non-semisimple pair that has the multiplicative expression
\begin{equation}
\psi_{\mu\nu}^{\log}
=
y(\tau,\rho)\,
\psi_{\mu\nu}^{L}, \hspace{1cm}
y(\tau,\rho)
=
-i\tau-\ln\cosh\rho.
\label{eq:exact_log_graviton}
\end{equation}
A generalized-eigenvector structure appears under the action of the zero-mode generators on the pair as
\begin{subequations}
\begin{align}
L_0\psi^L
&=
2\psi^L,
&
\bar L_0\psi^L
&=
0,
\label{eq:zero_modes_left}
\\[1mm]
L_0\psi^{\log}
&=
2\psi^{\log}
+
\frac{1}{2}\psi^L,
&
\bar L_0\psi^{\log}
&=
\frac{1}{2}\psi^L.
\label{eq:zero_modes_log_GJ}
\end{align}
\end{subequations}
such that the total zero-mode operator
\begin{equation}
D=L_0+\bar L_0,
\end{equation}
acts at the primary level according to
\begin{equation}
D\psi_{\mu\nu}^{L}
=
2\psi_{\mu\nu}^{L},
\qquad
D\psi_{\mu\nu}^{\log}
=
2\psi_{\mu\nu}^{\log}
+
\psi_{\mu\nu}^{L}.
\label{eq:intro_primary_jordan}
\end{equation}
The logarithmic pair therefore forms a rank-two Jordan block under
the total zero-mode action. 

The question addressed in the present work is not whether
the logarithmic sector possesses a Jordan structure, but how that
structure is encoded in the finite action of the two commuting
conformal zero modes. We introduce independent additive parameters
\(s\) and \(\bar s\) and consider the Virasoro zero-mode operator
\begin{equation}
U(s,\bar s)
=
e^{-sL_0-\bar s\bar L_0}.
\label{eq:intro_zero_mode_operator}
\end{equation}
In the bulk-inherited normalization, the individual zero modes
restrict to the logarithmic subspace as
\begin{equation}
L_0
=
h\mathbf 1+\frac{1}{2}N,
\qquad
\bar L_0
=
\bar h\mathbf 1+\frac{1}{2}N,
\qquad
N^2=0,
\label{eq:intro_zero_mode_jordan}
\end{equation}
where
\begin{equation}
N\psi^L=0,
\qquad
N\psi^{\log}=\psi^L,
\end{equation}
and \((h,\bar h)=(2,0)\) for critical TMG. Since the two zero modes
commute, their finite action is
\begin{equation}
U(s,\bar s)
=
e^{-hs-\bar h\bar s}
\left[
\mathbf 1-\frac{1}{2}(s+\bar s)N
\right].
\label{eq:intro_finite_zero_mode_action}
\end{equation}
The semisimple components determine the ordinary conformal grading,
whereas the nilpotent mixing is governed by
\(\frac{1}{2}(s+\bar s)\). The significance of this construction is
not the exponentiation of a Jordan block by itself, but the use of
the finite Virasoro zero-mode action to organize several structures
associated with the logarithmic representation.

A related logarithmic structure occurs directly in the bulk
gravitational solution. The relation between bulk monodromy, the
rank-two logarithmic sector, and logarithmic two-point functions was
studied in Ref.~\cite{Mvondo-She:2026egr}. The present work develops
a complementary representation-theoretic perspective based on the
finite action generated by \(L_0\) and \(\bar L_0\). In particular,
we keep the Virasoro zero-mode flow distinct from the logarithmic
branch structure of the generalized bulk solution. Although both
reflect the same underlying indecomposable logarithmic sector, they
arise from different analytic operations and should not be
identified.

We develop this framework through two applications. The first concerns
the logarithmic two-point functions. Introducing the dimensionless
boundary variables
\begin{equation}
s
=
\log\left(
\frac{z_{12}}{z_0}
\right),
\qquad
\bar s
=
\log\left(
\frac{\bar z_{12}}{\bar z_0}
\right),
\label{eq:intro_boundary_flow_coordinates}
\end{equation}
the semisimple part of the zero-mode action generates the ordinary
conformal power law, while the nilpotent mixing depends on the full
two-dimensional separation through
\begin{equation}
\frac{s+\bar s}{2}
=
\frac{1}{2}
\log\left(
\frac{z_{12}\bar z_{12}}
{z_0\bar z_0}
\right).
\label{eq:intro_boundary_nilpotent_coordinate}
\end{equation}
Together with the standard rank-two LCFT pairing conditions, this
yields the characteristic logarithmic coordinate dependence. For
critical TMG, the ordinary conformal power law is holomorphic because
\((h,\bar h)=(2,0)\), whereas the logarithmic term depends on the full
Euclidean separation. This reflects the fact that \(\bar L_0\) has
vanishing semisimple weight but retains a nontrivial nilpotent action
on the logarithmic partner.

The logarithmic two-point functions of critical TMG have previously
been derived directly from the gravitational theory using holographic
renormalization \cite{Skenderis:2009nt} and explicit bulk correlator
calculations \cite{Grumiller:2009mw}. The purpose of the present
analysis is not to rederive their gravitational normalization, but to
show how their universal logarithmic coordinate dependence can be
traced to the finite Virasoro zero-mode representation of the
logarithmic pair.

The second application concerns the logarithmic contribution to the
one-loop partition function. Its gravitational origin through the
one-loop determinant is already established
\cite{Gaberdiel:2010xv}; here we instead relate its conventional multiplicative grading variables to the additive zero-mode parameters by
\begin{equation}
q=e^{-s},
\qquad
\bar q=e^{-\bar s},
\label{eq:intro_q_from_s}
\end{equation}
so that
\begin{equation}
q^{L_0}\bar q^{\,\bar L_0}
=
e^{-sL_0-\bar s\bar L_0}
=
U(s,\bar s).
\label{eq:intro_flow_grading_identity}
\end{equation}
This expresses the familiar conformal state-counting variables in
terms of the additive parameters of the same zero-mode action. The
resulting formulation is an exact reparameterization of the
established logarithmic partition function and does not modify its
spectrum, multiplicities, descendant structure, or multiparticle
interpretation.

These two applications probe complementary aspects of the same
indecomposable representation. Acting on an individual logarithmic
state, \(U(s,\bar s)\) retains the off-diagonal contribution
proportional to
\(\frac{1}{2}(s+\bar s)N\), and the corresponding Jordan mixing is
visible in logarithmic correlation functions. Inside an ordinary
graded trace, the nilpotent matrix element is not separately
resolved; the trace instead organizes conformal weights,
multiplicities, and descendant content. The relation between
correlators and partition-function grading therefore lies in their
common representation-theoretic origin, rather than in an
identification of the observables themselves.

The contribution of the present work is thus to place the logarithmic
two-point functions and the established partition-function grading
within a common Virasoro zero-mode framework. The previously established
combinatorial, geometric, and integrable interpretations of the
logarithmic partition function
\cite{Mvondo-She:2018htn,Mvondo-She:2019vbx,
Mvondo-She:2021joh,Mvondo-She:2022jnf}
remain unchanged.

The paper is organized as follows.
Section~\ref{sec:virasoro_flow} develops the Virasoro zero-mode flow
and its finite action on the logarithmic pair, and clarifies its
distinction from the logarithmic branch structure of the bulk theory.
Section~\ref{sec:two_point_functions} applies the zero-mode action to
the rank-two logarithmic two-point functions and studies their
analytic continuation.
Section~\ref{sec:flow_partition_function} relates the additive
zero-mode parameters to the conventional conformal grading of the
logarithmic partition function and distinguishes finite Jordan
evolution from graded state counting.
Finally, Section~\ref{sec:sum_and_out} summarizes the main results
and discusses possible extensions.


\section{Virasoro zero-mode flow and logarithmic structure}
\label{sec:virasoro_flow}

The logarithmic sector of critical Topologically Massive Gravity is
characterized by a non-diagonalizable action of the conformal zero
modes on the logarithmic pair. For the logarithmic graviton, this
Jordan structure is not confined to the action of \(L_0\) alone:
both \(L_0\) and \(\bar L_0\) act non-diagonalizably on the
degenerate left-moving and logarithmic states. It is therefore
natural to consider the finite Virasoro zero-mode action generated
jointly by the two commuting conformal zero modes. We accordingly introduce independent additive parameters \(s\) and \(\bar s\) and consider the finite transformation generated by \(L_0\) and \(\bar L_0\). 

\subsection{Virasoro flow generated by the conformal zero modes}
\label{subsec:complex_zero_mode_flow}

Let \(L_0\) and \(\bar L_0\) act on the relevant state space
\(\mathcal H\). Since the two conformal zero modes commute,
\begin{equation}
[L_0,\bar L_0]=0,
\label{eq:zero_mode_commutator}
\end{equation}
we define the Virasoro zero-mode flow by
\begin{equation}
U(s,\bar s)
=
e^{-sL_0-\bar s\bar L_0}
=
e^{-sL_0}e^{-\bar s\bar L_0},
\qquad
(s,\bar s)\in\mathbb C^2.
\label{eq:complex_flow_operator}
\end{equation}
For any \(\psi\in\mathcal H\), its finite transform is
\begin{equation}
\psi(s,\bar s)
=
U(s,\bar s)\psi.
\label{eq:complex_flow_state}
\end{equation}
Differentiation with respect to the two additive parameters gives
\begin{subequations}
\begin{align}
\frac{\partial}{\partial s}\psi(s,\bar s)
&=
-L_0\psi(s,\bar s),
\label{eq:complex_flow_equation_s}
\\
\frac{\partial}{\partial\bar s}\psi(s,\bar s)
&=
-\bar L_0\psi(s,\bar s),
\label{eq:complex_flow_equation_sbar}
\end{align}
\end{subequations}
with initial condition
\(\psi(0,0)=\psi\). The two evolution equations are compatible
because
\begin{subequations}
\begin{align}
\frac{\partial^2}{\partial s\,\partial\bar s}
\psi(s,\bar s)
&=
L_0\bar L_0\,\psi(s,\bar s),
\\
\frac{\partial^2}{\partial\bar s\,\partial s}
\psi(s,\bar s)
&=
\bar L_0L_0\,\psi(s,\bar s),
\end{align}
\end{subequations}
and hence
\begin{equation}
\left(
\frac{\partial^2}{\partial s\,\partial\bar s}
-
\frac{\partial^2}{\partial\bar s\,\partial s}
\right)
\psi(s,\bar s)
=
[L_0,\bar L_0]\psi(s,\bar s)
=
0.
\label{eq:flow_compatibility}
\end{equation}
The finite transformations also satisfy the additive composition law
\begin{equation}
U(s_1,\bar s_1)\,
U(s_2,\bar s_2)
=
U(s_1+s_2,\bar s_1+\bar s_2).
\label{eq:flow_composition}
\end{equation}

For a simultaneous eigenstate of conformal weights
\((h,\bar h)\),
\begin{equation}
L_0\psi=h\psi,
\qquad
\bar L_0\psi=\bar h\psi,
\label{eq:diagonal_zero_mode_action}
\end{equation}
the finite zero-mode action is purely multiplicative,
\begin{equation}
U(s,\bar s)\psi
=
e^{-hs-\bar h\bar s}\psi.
\label{eq:diagonal_flow}
\end{equation}
Thus no state mixing occurs on a simultaneous eigenspace. The
logarithmic sector differs because the zero-mode action is
non-diagonalizable and contains a nilpotent component. In critical
TMG this structure involves both \(L_0\) and \(\bar L_0\). As shown
in the next subsection, their action on the logarithmic pair contains
equal nilpotent contributions, and exponentiation separates the
semisimple conformal weights from the resulting nilpotent mixing.

At this stage, \(s\) and \(\bar s\) are independent complex
parameters of the Virasoro zero-mode flow, with no intrinsic identification with bulk spacetime coordinates.

\subsection{Finite zero-mode flow on the logarithmic pair}
\label{subsec:finite_flow_logarithmic_pair}

We now specialize the Virasoro zero-mode flow to the logarithmic pair
of critical TMG. 

Substituting Eq.~\eqref{eq:intro_zero_mode_jordan} into the Virasoro
zero-mode flow defined in Eq.~\eqref{eq:complex_flow_operator} gives
\begin{equation}
U(s,\bar s)
=
e^{-sL_0-\bar s\bar L_0}
=
e^{-hs-\bar h\bar s}
e^{-\frac{1}{2}(s+\bar s)N}
=
e^{-hs-\bar h\bar s}
\left[
\mathbf 1-\frac{1}{2}(s+\bar s)N
\right],
\label{eq:finite_complex_flow}
\end{equation}
where the nilpotent exponential truncates exactly because \(N^2=0\).
Consequently,
\begin{subequations}
\begin{align}
U(s,\bar s)\psi^L
&=
e^{-hs-\bar h\bar s}\psi^L,
\label{eq:flow_primary}
\\[1mm]
U(s,\bar s)\psi^{\log}
&=
e^{-hs-\bar h\bar s}
\left[
\psi^{\log}
-
\frac{1}{2}(s+\bar s)\psi^L
\right].
\label{eq:flow_logarithmic}
\end{align}
\end{subequations}

Equation~\eqref{eq:flow_logarithmic} gives the finite Virasoro
zero-mode action on the logarithmic partner. The semisimple
components of \(L_0\) and \(\bar L_0\) generate the ordinary
conformal factor \(e^{-hs-\bar h\bar s}\), whereas their equal
nilpotent contributions combine to produce the mixing governed by
\(\frac{1}{2}(s+\bar s)\). In particular, for
critical TMG the antiholomorphic zero mode contributes to the
nilpotent mixing despite its vanishing semisimple weight
\(\bar h=0\).

\subsection{Virasoro zero-mode flow versus bulk logarithmic monodromy}
\label{subsec:continuous_flow_monodromy}

The Virasoro zero-mode flow derived in the preceding subsection
describes the finite representation-theoretic action generated by
the holomorphic and antiholomorphic conformal zero modes. On the
logarithmic pair, the nilpotent mixing is governed by
\(\frac{1}{2}(s+\bar s)\), as follows from
Eq.~\eqref{eq:finite_complex_flow}. This conformal action should be
distinguished from the logarithmic branch structure carried by the
generalized bulk solution. The latter originates from the explicit
spacetime dependence of the logarithmic gravitational mode and has
been studied from the viewpoint of bulk monodromy in
Ref.~\cite{Mvondo-She:2026egr}. Although both structures reflect the
same underlying indecomposable logarithmic representation, they
correspond to different analytic operations.

The distinction is already visible at the level of the conformal
zero modes. Consider a simultaneous continuation of the two flow
parameters in opposite directions,
\begin{equation}
(s,\bar s)
\longrightarrow
\left(
s+2\pi i n,
\bar s-2\pi i n
\right),
\qquad
n\in\mathbb Z.
\label{eq:joint_flow_opposite_shift}
\end{equation}
The combination governing the nilpotent mixing is invariant,
\begin{equation}
s+\bar s
\longrightarrow
s+\bar s.
\label{eq:joint_flow_nilpotent_invariant}
\end{equation}
Equivalently, the corresponding finite zero-mode operator is
\begin{equation}
U(2\pi i n,-2\pi i n)
=
\exp\left[
-2\pi i n
\left(
L_0-\bar L_0
\right)
\right]
=
e^{-2\pi i n(h-\bar h)}\mathbf 1,
\label{eq:joint_flow_opposite_operator}
\end{equation}
because the equal nilpotent contributions \(\frac12N\) in
\(L_0\) and \(\bar L_0\) cancel in their difference. For the
critical-TMG weights $(h,\bar h)=(2,0)$, the remaining semisimple factor is unity for integer \(n\).

Thus the Virasoro zero-mode flow does not generate a nontrivial
nilpotent monodromy under the opposite continuation
\eqref{eq:joint_flow_opposite_shift}. This differs from bulk
logarithmic monodromy, which arises from analytic continuation of the
generalized gravitational solution itself. The two constructions
therefore reflect the same underlying logarithmic structure while
having distinct analytic origins and transformation laws.


\section{Logarithmic two-point functions from the Virasoro zero-mode flow}
\label{sec:two_point_functions}

We now apply the Virasoro zero-mode flow developed in
Section~\ref{sec:virasoro_flow} to the two-point functions of the
critical-TMG logarithmic pair. The aim is to determine how the finite
zero-mode action constrains their coordinate dependence and, in
particular, how the nilpotent mixing generates the logarithmic term
characteristic of a rank-two LCFT.

Starting from the finite transformation of
\(\psi^L\) and \(\psi^{\log}\) derived above, we first obtain the
flow equations satisfied by the two-point functions. The standard
rank-two LCFT pairing conditions then reduce these equations to the
mixed and logarithmic correlators appropriate to critical TMG. This
derivation isolates the representation-theoretic origin of the
logarithmic coordinate dependence; its relation to the independently
defined bulk logarithmic structure will be considered after the
correlators have been obtained.

\subsection{Flow of the logarithmic pair}
\label{subsec:flow_log_pair_correlators}

By the state--operator correspondence, we associate the asymptotic
states \(\psi^L\) and \(\psi^{\log}\) with the corresponding boundary fields, denoted by the same symbols. To express the finite zero-mode transformations derived in Section~\ref{sec:virasoro_flow} in terms of the coordinate dependence of two-point functions, introduce the dimensionless
separations
\begin{equation}
\zeta
=
\frac{z_{12}}{z_0},
\qquad
\bar\zeta
=
\frac{\bar z_{12}}{\bar z_0},
\qquad
z_{12}=z_1-z_2,
\label{eq:dimensionless_separations}
\end{equation}
where \(z_0\) and \(\bar z_0\) are reference scales, and identify the
additive flow parameters as
\begin{equation}
s=\log\zeta,
\qquad
\bar s=\log\bar\zeta,
\label{eq:flow_coordinate_separations}
\end{equation}
or equivalently,
\begin{equation}
\zeta=e^s,
\qquad
\bar\zeta=e^{\bar s}.
\label{eq:zeta_exp_flow}
\end{equation}

For a two-point function, the semisimple contribution from the two
field insertions is therefore
\begin{equation}
e^{-2hs-2\bar h\bar s}
=
\zeta^{-2h}\bar\zeta^{-2\bar h}
=
\left(
\frac{z_0}{z_{12}}
\right)^{2h}
\left(
\frac{\bar z_0}{\bar z_{12}}
\right)^{2\bar h}.
\label{eq:scaling_factor_two_point}
\end{equation}
The parameter combination governing the nilpotent mixing becomes
\begin{equation}
s+\bar s
=
\log\left(
\frac{z_{12}}{z_0}
\right)
+
\log\left(
\frac{\bar z_{12}}{\bar z_0}
\right)
=
\log\left(
\frac{z_{12}\bar z_{12}}
{z_0\bar z_0}
\right),
\label{eq:nilpotent_coordinate_combination}
\end{equation}
up to the usual choice of logarithmic branches. Upon imposing the Euclidean conjugacy conditions \(\bar z_{12}=z_{12}^{*}\) and
\(\bar z_0=z_0^{*}\), this reduces to
\begin{equation}
s+\bar s
=
\log\left|
\frac{z_{12}}{z_0}
\right|^2.
\label{eq:euclidean_logarithmic_separation}
\end{equation}

The zero-mode flow thus separates the two ingredients entering the
logarithmic correlators: the semisimple action determines the ordinary
conformal power law, while the nilpotent mixing introduces dependence
on the logarithm of the full two-dimensional separation.
\subsection{Primary and mixed two-point functions}
\label{subsec:primary_mixed_correlators}

Let
\begin{equation}
A
=
\left\langle
\psi^L(z_0,\bar z_0)\psi^L(0,0)
\right\rangle
\label{eq:A_definition}
\end{equation}
denote the primary--primary pairing at the reference separation.
The finite zero-mode action in
Eq.~\eqref{eq:finite_complex_flow} gives
\begin{equation}
\left\langle
\psi^L(z_1,\bar z_1)\psi^L(z_2,\bar z_2)
\right\rangle
=
e^{-2hs-2\bar h\bar s}A
=
A
\left(
\frac{z_0}{z_{12}}
\right)^{2h}
\left(
\frac{\bar z_0}{\bar z_{12}}
\right)^{2\bar h}.
\label{eq:LL_general}
\end{equation}
The Virasoro zero-mode flow fixes the coordinate dependence but not the
constant \(A\). For the standard rank-two logarithmic pairing, the
primary member is null,
\begin{equation}
A=0,
\label{eq:A_zero}
\end{equation}
and hence
\begin{equation}
\left\langle
\psi^L(z_1,\bar z_1)\psi^L(z_2,\bar z_2)
\right\rangle
=
0.
\label{eq:LL_final}
\end{equation}

For the mixed pairings, define
\begin{equation}
b
=
\left\langle
\psi^L(z_0,\bar z_0)\psi^{\log}(0,0)
\right\rangle,
\qquad
\widetilde b
=
\left\langle
\psi^{\log}(z_0,\bar z_0)\psi^L(0,0)
\right\rangle.
\label{eq:mixed_reference_constants}
\end{equation}
For convenience, introduce the flow-parameter combination
\begin{equation}
\sigma
\equiv
s+\bar s.
\label{eq:sigma_definition}
\end{equation}
Using the finite logarithmic mixing encoded in
Eq.~\eqref{eq:finite_complex_flow}, the mixed two-point functions
take the form
\begin{align}
\left\langle
\psi^L(z_1,\bar z_1)
\psi^{\log}(z_2,\bar z_2)
\right\rangle
&=
e^{-2hs-2\bar h\bar s}
\left(
b-\frac{\sigma}{2}A
\right),
\label{eq:Llog_general_flow}
\\[2mm]
\left\langle
\psi^{\log}(z_1,\bar z_1)
\psi^L(z_2,\bar z_2)
\right\rangle
&=
e^{-2hs-2\bar h\bar s}
\left(
\widetilde b-\frac{\sigma}{2}A
\right).
\label{eq:logL_general_flow}
\end{align}
The terms proportional to \(\sigma A/2\) arise from the nilpotent
mixing in the zero-mode action. Imposing the logarithmic pairing
condition \(A=0\) and using
Eq.~\eqref{eq:scaling_factor_two_point} gives
\begin{align}
\left\langle
\psi^L(z_1,\bar z_1)
\psi^{\log}(z_2,\bar z_2)
\right\rangle
&=
b
\left(
\frac{z_0}{z_{12}}
\right)^{2h}
\left(
\frac{\bar z_0}{\bar z_{12}}
\right)^{2\bar h},
\label{eq:Llog_final_general_b}
\\[2mm]
\left\langle
\psi^{\log}(z_1,\bar z_1)
\psi^L(z_2,\bar z_2)
\right\rangle
&=
\widetilde b
\left(
\frac{z_0}{z_{12}}
\right)^{2h}
\left(
\frac{\bar z_0}{\bar z_{12}}
\right)^{2\bar h}.
\label{eq:logL_final_general_b}
\end{align}
For a symmetric two-point pairing,
\begin{equation}
\widetilde b=b,
\label{eq:b_symmetry}
\end{equation}
and therefore
\begin{equation}
\left\langle
\psi^L(z_1,\bar z_1)
\psi^{\log}(z_2,\bar z_2)
\right\rangle
=
\left\langle
\psi^{\log}(z_1,\bar z_1)
\psi^L(z_2,\bar z_2)
\right\rangle
=
b
\left(
\frac{z_0}{z_{12}}
\right)^{2h}
\left(
\frac{\bar z_0}{\bar z_{12}}
\right)^{2\bar h}.
\label{eq:mixed_final}
\end{equation}
For the critical-TMG logarithmic pair,
\((h,\bar h)=(2,0)\), so that the mixed correlator reduces to
\begin{equation}
\left\langle
\psi^L(z_1,\bar z_1)
\psi^{\log}(z_2,\bar z_2)
\right\rangle
=
b
\left(
\frac{z_0}{z_{12}}
\right)^4.
\label{eq:mixed_final_tmg}
\end{equation}
Thus the non-diagonalizable \(\bar L_0\) action does not introduce an
antiholomorphic power law when \(\bar h=0\), although its nilpotent
contribution remains essential for the logarithmic dependence of the
logarithmic--logarithmic correlator. The constant \(b\) depends on the
normalization of the logarithmic pair and is not determined by the
Virasoro zero-mode flow.

\subsection{The logarithmic two-point function}
\label{subsec:log_log_correlator}

Let
\begin{equation}
d
=
\left\langle
\psi^{\log}(z_0,\bar z_0)
\psi^{\log}(0,0)
\right\rangle
\label{eq:d_reference}
\end{equation}
denote the logarithmic--logarithmic pairing at the reference
separation. Applying the finite zero-mode action
Eq.~\eqref{eq:finite_complex_flow} to both insertions gives
\begin{subequations}
\begin{align}
\left\langle
\psi^{\log}(z_1,\bar z_1)
\psi^{\log}(z_2,\bar z_2)
\right\rangle
&=
e^{-2hs-2\bar h\bar s}
\left\langle
\left(
\psi^{\log}-\frac{\sigma}{2}\psi^L
\right)
\left(
\psi^{\log}-\frac{\sigma}{2}\psi^L
\right)
\right\rangle
\\[1mm]
&=
e^{-2hs-2\bar h\bar s}
\left[
d
-\frac{\sigma}{2}\left(b+\widetilde b\right)
+\frac{\sigma^2}{4}A
\right],
\label{eq:loglog_general_flow}
\end{align}
\end{subequations}
where \(\sigma=s+\bar s\) is the flow-parameter combination introduced
in Eq.~\eqref{eq:sigma_definition}. The linear and quadratic terms in
\(\sigma\) arise from one and two nilpotent contributions,
respectively.

For the standard rank-two logarithmic pairing,
\begin{equation}
A=0,
\qquad
\widetilde b=b,
\label{eq:standard_log_pairing}
\end{equation}
and hence
\begin{equation}
\left\langle
\psi^{\log}(z_1,\bar z_1)
\psi^{\log}(z_2,\bar z_2)
\right\rangle
=
e^{-2hs-2\bar h\bar s}
\left(
d-b\sigma
\right).
\label{eq:loglog_sigma_form}
\end{equation}
Using
Eqs.~\eqref{eq:flow_coordinate_separations},
\eqref{eq:scaling_factor_two_point}, and
\eqref{eq:nilpotent_coordinate_combination}, this becomes
\begin{equation}
\left\langle
\psi^{\log}(z_1,\bar z_1)
\psi^{\log}(z_2,\bar z_2)
\right\rangle
=
\left(
\frac{z_0}{z_{12}}
\right)^{2h}
\left(
\frac{\bar z_0}{\bar z_{12}}
\right)^{2\bar h}
\left[
d
-
b
\log\left(
\frac{z_{12}\bar z_{12}}
{z_0\bar z_0}
\right)
\right].
\label{eq:loglog_final}
\end{equation}
Under Euclidean conjugacy conditions,
\(\bar z_{12}=z_{12}^{*}\) and
\(\bar z_0=z_0^{*}\), so that
\begin{equation}
\left\langle
\psi^{\log}(z_1,\bar z_1)
\psi^{\log}(z_2,\bar z_2)
\right\rangle
=
\left(
\frac{z_0}{z_{12}}
\right)^{2h}
\left(
\frac{\bar z_0}{\bar z_{12}}
\right)^{2\bar h}
\left[
d
-
b
\log
\left|
\frac{z_{12}}{z_0}
\right|^2
\right].
\label{eq:loglog_euclidean}
\end{equation}
For the critical-TMG logarithmic pair,
\((h,\bar h)=(2,0)\), this reduces to
\begin{equation}
\left\langle
\psi^{\log}(z_1,\bar z_1)
\psi^{\log}(z_2,\bar z_2)
\right\rangle
=
\left(
\frac{z_0}{z_{12}}
\right)^4
\left[
d
-
b
\log
\left|
\frac{z_{12}}{z_0}
\right|^2
\right].
\label{eq:loglog_tmg}
\end{equation}
The appearance of the full logarithmic separation
\(\log |z_{12}/z_0|^2\) follows from the equal nilpotent
contributions of the holomorphic and antiholomorphic zero modes.
Thus, although \(\bar h=0\) for the critical-TMG logarithmic primary,
the antiholomorphic zero mode contributes nontrivially through its
nilpotent part.

The additive constant \(d\) is basis dependent. Under the field
redefinition
\begin{equation}
\psi^{\log}
\longrightarrow
\psi^{\log}
+
\alpha\psi^L,
\label{eq:log_field_redefinition}
\end{equation}
the zero-mode Jordan relations are preserved. At the level of the
reference pairings,
\begin{equation}
d
\longrightarrow
d
+
\alpha\left(b+\widetilde b\right)
+
\alpha^2 A.
\label{eq:d_shift_general}
\end{equation}
After imposing the standard logarithmic pairing conditions
\(A=0\) and \(\widetilde b=b\), this reduces to
\begin{equation}
d
\longrightarrow
d+2\alpha b.
\label{eq:d_shift}
\end{equation}
The additive constant \(d\) is therefore basis dependent, whereas
the coefficient of the logarithmic term is fixed relative to the
mixed pairing once the normalization of the logarithmic partner has
been chosen. The Virasoro zero-mode flow determines this relation
between the mixed and logarithmic--logarithmic coordinate dependence,
while the overall normalization \(b\) remains independent of the
flow construction.

The critical-TMG correlator is therefore chiral in its ordinary
conformal weights but not purely holomorphic in its logarithmic
dependence. This distinction reflects the indecomposable action of
both conformal zero modes on the logarithmic partner.

\subsection{Analytic continuation and the logarithmic correlator}
\label{subsec:correlator_monodromy}

We now examine the analytic continuation of the logarithmic
two-point function in the complexified boundary coordinates. Before
restricting to the Euclidean conjugate locus, the logarithmic
coordinate appearing in the two-point function is
\begin{equation}
\sigma
=
s+\bar s
=
\log\left(
\frac{z_{12}}{z_0}
\right)
+
\log\left(
\frac{\bar z_{12}}{\bar z_0}
\right).
\label{eq:sigma_correlator_continuation}
\end{equation}
The two logarithms may be analytically continued independently. If
\begin{equation}
z_{12}
\longrightarrow
e^{2\pi i n}z_{12},
\qquad
\bar z_{12}\ \text{fixed},
\qquad
n\in\mathbb Z,
\label{eq:holomorphic_continuation_correlator}
\end{equation}
then
\begin{equation}
s
\longrightarrow
s+2\pi i n,
\qquad
\bar s
\longrightarrow
\bar s,
\label{eq:holomorphic_flow_shift_correlator}
\end{equation}
and therefore
\begin{equation}
\sigma
\longrightarrow
\sigma+2\pi i n.
\label{eq:sigma_holomorphic_shift}
\end{equation}
The logarithmic factor consequently acquires the branch shift
\begin{equation}
d-b\sigma
\longrightarrow
d-b\sigma-2\pi i n\,b.
\label{eq:loglog_holomorphic_branch_shift}
\end{equation}
For the critical-TMG weight \(h=2\), the semisimple factor
\(z_{12}^{-2h}=z_{12}^{-4}\) is single-valued under this winding, so
the nontrivial branch transformation is entirely contained in the
logarithmic factor.

The Euclidean angular continuation is different. Under Euclidean conjugacy condition, \(\bar z_{12}=z_{12}^{*}\), and an angular winding acts as
\begin{equation}
z_{12}
\longrightarrow
e^{2\pi i n}z_{12},
\qquad
\bar z_{12}
\longrightarrow
e^{-2\pi i n}\bar z_{12}.
\label{eq:euclidean_angular_continuation_correlator}
\end{equation}
Equivalently,
\begin{equation}
(s,\bar s)
\longrightarrow
\left(
s+2\pi i n,
\bar s-2\pi i n
\right).
\label{eq:euclidean_flow_shift_correlator}
\end{equation}
Hence
\begin{equation}
\sigma=s+\bar s
\longrightarrow
\sigma,
\label{eq:sigma_euclidean_invariant}
\end{equation}
and therefore
\begin{equation}
\log\left|
\frac{z_{12}}{z_0}
\right|^2
\longrightarrow
\log\left|
\frac{z_{12}}{z_0}
\right|^2.
\label{eq:euclidean_log_invariant}
\end{equation}
The Euclidean logarithmic two-point function is thus single-valued
under simultaneous winding of the conjugate coordinates. This is
consistent with the Virasoro zero-mode action discussed in
Section~\ref{subsec:continuous_flow_monodromy}, since the relevant
generator is \(L_0-\bar L_0\), for which the equal nilpotent
contributions cancel.

\subsection{Comparison with holographic two-point functions}
\label{subsec:comparison_holographic_correlators}

The logarithmic two-point functions obtained above from the Virasoro
zero-mode flow reproduce the characteristic coordinate dependence
previously derived directly from the bulk theory by holographic
methods. In particular, Skenderis, Taylor and van Rees
\cite{Skenderis:2009nt} developed the holographic renormalization of
topologically massive gravity and computed the one- and two-point
functions at the chiral point. In the standard LCFT normalization,
their logarithmic pair \((T,t)\) satisfies
\begin{equation}
\langle T(z)T(0)\rangle=0,
\qquad
\langle T(z)t(0)\rangle
=
\frac{b_{\rm LCFT}}{2z^4},
\label{eq:holographic_Tt}
\end{equation}
and
\begin{equation}
\langle t(z,\bar z)t(0)\rangle
=
-\frac{b_{\rm LCFT}}{z^4}
\log\!\left(m^2|z|^2\right),
\label{eq:holographic_tt}
\end{equation}
up to the usual freedom to redefine the logarithmic partner by
\(t\rightarrow t+\alpha T\). For critical TMG the holographic
calculation gives
\begin{equation}
c_L=0,
\qquad
c_R=\frac{3\ell}{G_N},
\qquad
b_{\rm LCFT}=-\frac{3\ell}{G_N},
\label{eq:holographic_new_anomaly}
\end{equation}
where the factors of the AdS radius depend on the convention in which
\(\ell\) is restored.

The same logarithmic structure was subsequently obtained by Grumiller
and Sachs \cite{Grumiller:2009mw} from an explicit gravity-side
calculation employing regular non-normalizable left, right and
logarithmic modes in global \(\mathrm{AdS}_3\). Their results likewise
reproduce the LCFT two-point functions and the corresponding new
anomaly.

These holographic calculations provide a direct comparison with the
present construction. Here the functional form of the correlators is
not obtained by evaluating and renormalizing the bulk on-shell action,
but can be traced to the finite Virasoro zero-mode action on the logarithmic pair, together with the standard rank-two LCFT pairing conditions. After imposing the standard rank-two LCFT pairing
conditions, the zero-mode flow gives
\begin{equation}
\langle \psi^L(z,\bar z)\psi^L(0)\rangle=0,
\label{eq:flow_LL_comparison}
\end{equation}
\begin{equation}
\langle \psi^L(z,\bar z)\psi^{\log}(0)\rangle
=
\frac{b_{\rm f}}
{z^{2h}\bar z^{2\bar h}},
\label{eq:flow_Llog_comparison}
\end{equation}
and
\begin{equation}
\langle \psi^{\log}(z,\bar z)\psi^{\log}(0)\rangle
=
\frac{
d_{\rm f}
-b_{\rm f}\log|z/z_0|^2
}
{z^{2h}\bar z^{2\bar h}},
\label{eq:flow_loglog_comparison}
\end{equation}
where \(b_{\rm f}\) and \(d_{\rm f}\) denote the constants in the
normalization adopted in the present zero-mode construction. At the
chiral point, \((h,\bar h)=(2,0)\), these expressions have the same
rank-two LCFT coordinate structure as the holographic correlators in
Refs.~\cite{Skenderis:2009nt,Grumiller:2009mw}.

The normalization dictionary must take into account the Jordan
normalization as well as the two-point-function normalization.
In the convention used here,
\begin{equation}
L_0\psi^{\log}
=
h\psi^{\log}
+
\frac{1}{2}\psi^L,
\end{equation}
whereas the standard LCFT logarithmic partner \(t\) is conventionally
normalized with unit Jordan coefficient. Identifying
\(T=\psi^L\) therefore requires
\begin{equation}
t=2\psi^{\log}.
\label{eq:canonical_log_partner_rescaling}
\end{equation}
Consequently,
\begin{equation}
\langle T(z)t(0)\rangle
=
2\langle\psi^L(z)\psi^{\log}(0)\rangle
=
\frac{2b_f}{z^4}.
\end{equation}
Comparison with the standard holographic normalization
\begin{equation}
\langle T(z)t(0)\rangle
=
\frac{b_{\mathrm{LCFT}}}{2z^4}
\end{equation}
therefore gives
\begin{equation}
b_{\mathrm{LCFT}}
=
4b_{\rm f}.
\label{eq:bf_blcft_dictionary}
\end{equation}
The logarithmic--logarithmic correlator provides the same
identification
\begin{equation}
\langle t(z,\bar z)t(0)\rangle
=
4\langle\psi^{\log}(z,\bar z)
        \psi^{\log}(0)\rangle
=
\frac{
4d_{\rm f}
-
4b_{\rm f}\log|z/z_0|^2
}{z^4}.
\end{equation}
Thus the coefficient of the logarithmic term becomes
\(-b_{\mathrm{LCFT}}\), in agreement with the standard LCFT
normalization after using
Eq.~\eqref{eq:bf_blcft_dictionary}.

The additive constant in the logarithmic--logarithmic correlator has
the same basis-dependent status in the two descriptions. In the
present convention,
\begin{equation}
\psi^{\log}
\longrightarrow
\psi^{\log}+\alpha\psi^L
\end{equation}
induces
\begin{equation}
d_{\rm f}
\longrightarrow
d_{\rm f}+2\alpha b_{\rm f}.
\label{eq:df_shift_comparison}
\end{equation}
Likewise, the non-logarithmic constant in the general rank-two LCFT
logarithmic--logarithmic correlator may be shifted by a redefinition
of the logarithmic partner. Thus \(d_{\rm f}\) has the same basis-dependent
status as the corresponding additive constant in the standard LCFT
correlator.

The comparison also clarifies the scope of the zero-mode derivation.
The Virasoro zero-mode flow fixes the universal coordinate dependence
and the logarithmic mixing dictated by the Jordan representation, but
does not by itself determine the dynamical normalization of the
two-point functions. In critical TMG that normalization is fixed by
the bulk gravitational theory; in the holographic convention of
Ref.~\cite{Skenderis:2009nt}, it is encoded in the new anomaly
\(b_{\rm LCFT}=-3\ell/G_N\). The zero-mode construction and the
holographic calculation are therefore complementary: the former
isolates the representation-theoretic origin of the logarithmic
dependence, whereas the latter determines its gravitational
normalization.
\section{Virasoro zero-mode flow and the logarithmic partition function}
\label{sec:flow_partition_function}

We now relate the additive parameters \((s,\bar s)\) of the Virasoro zero-mode flow to the multiplicative grading variables \((q,\bar q)\) appearing in the logarithmic contribution to the partition function of critical TMG. This places its conventional conformal grading in the framework developed in the preceding sections, provides a representation-theoretic connection between the finite action on the logarithmic pair and the established state-counting description of the logarithmic sector.

\subsection{From additive flow parameters to conformal grading}
\label{subsec:additive_to_conformal_grading}

Introduce the multiplicative grading variables
\begin{equation}
q=e^{-s},
\qquad
\bar q=e^{-\bar s}.
\label{eq:q_flow_coordinates}
\end{equation}
Since \(L_0\) and \(\bar L_0\) commute, the corresponding joint
grading operator is
\begin{equation}
q^{L_0}\bar q^{\,\bar L_0}
=
e^{-sL_0}e^{-\bar s\bar L_0}
=
e^{-sL_0-\bar s\bar L_0}
=
U(s,\bar s).
\label{eq:qL0_qbarL0_flow_identity}
\end{equation}
Thus \((s,\bar s)\) and \((q,\bar q)\) provide additive and
multiplicative parameterizations, respectively, of the same Virasoro
zero-mode action.

Using the Jordan decomposition established in
Section~\ref{sec:virasoro_flow}, the action of the grading operator on
the logarithmic pair is
\begin{equation}
q^{L_0}\bar q^{\,\bar L_0}
=
e^{-hs-\bar h\bar s}
e^{-\frac{1}{2}(s+\bar s)N}
=
q^h\bar q^{\,\bar h}
\left[
\mathbf 1-\frac{1}{2}(s+\bar s)N
\right].
\label{eq:log_joint_grading_partition}
\end{equation}
Equivalently, using
\(s=-\log q\) and \(\bar s=-\log\bar q\), this may be written as
\begin{equation}
q^{L_0}\bar q^{\,\bar L_0}
=
q^h\bar q^{\,\bar h}
\left[
\mathbf 1+\frac{1}{2}\log(q\bar q)\,N
\right],
\label{eq:log_joint_grading_q}
\end{equation}
with a consistent choice of logarithmic branches. The factor
\(q^h\bar q^{\,\bar h}\) is determined by the semisimple components
of the two zero modes, whereas the nilpotent mixing is governed by
the combination
\begin{equation}
\frac{s+\bar s}{2}
=
-\frac{1}{2}\log(q\bar q).
\label{eq:nilpotent_grading_combination}
\end{equation}
Equation~\eqref{eq:log_joint_grading_partition} describes the finite
action of the grading operator on the logarithmic subspace. It does
not imply that the ordinary graded trace separately resolves the
off-diagonal nilpotent matrix element. The distinction between the
state-space action and the information retained by the trace will be
examined in Section~\ref{subsec:jordan_evolution_state_counting}.

For the conventional Euclidean grading, the independent additive
parameters may be restricted to the conjugate locus
\begin{equation}
\bar s=s^{*}.
\label{eq:euclidean_grading_locus}
\end{equation}
In particular, we parameterize this locus by
\begin{equation}
s=u+i\theta,
\qquad
\bar s=u-i\theta,
\label{eq:s_u_theta_partition}
\end{equation}
with \(u,\theta\in\mathbb R\). The multiplicative variables then take
the form
\begin{equation}
q=e^{-u-i\theta},
\qquad
\bar q=e^{-u+i\theta},
\label{eq:q_u_theta_partition}
\end{equation}
so that
\begin{equation}
q\bar q=e^{-2u},
\qquad
\frac{s+\bar s}{2}
=
u
=
-\frac{1}{2}\log(q\bar q).
\label{eq:nilpotent_grading_u}
\end{equation}
The variable \(u\) therefore controls the nilpotent mixing on the
conjugate locus, while the phase \(\theta\) remains present in the
semisimple conformal grading whenever \(h\neq\bar h\). In particular,
for the critical-TMG weights \((h,\bar h)=(2,0)\),
\begin{equation}
q^h\bar q^{\,\bar h}
=
q^2
=
e^{-2u-2i\theta},
\label{eq:critical_tmg_semisimple_grading}
\end{equation}
whereas the coefficient of the nilpotent generator depends only on
\(u\). This distinction between semisimple grading and nilpotent
mixing will also be reflected in the partition function below.

The parameters \((u,\theta)\) introduced here are simply real
coordinates on the conjugate locus of the conformal grading
parameters. The construction in this section
is entirely determined by the representation-theoretic grading
operator \(e^{-sL_0-\bar s\bar L_0}\).

Equation~\eqref{eq:q_flow_coordinates} provides the change of
variables used in the next subsection to rewrite the established
logarithmic partition function in terms of the additive grading
parameters \((s,\bar s)\).

\subsection{The logarithmic partition function in additive grading coordinates}
\label{subsec:log_partition_additive_coordinates}

The logarithmic contribution to the one-loop partition function of
critical TMG may be written as \cite{Gaberdiel:2010xv}
\begin{equation}
\log Z_{\log}(q,\bar q)
=
\sum_{k=1}^{\infty}
\frac{q^{2k}}
{k\left(1-q^k\right)\left(1-\bar q^k\right)}.
\label{eq:log_partition_q}
\end{equation}
The factor \(q^{2k}\) carries the grading associated with the
left-moving logarithmic primary of weights
\((h,\bar h)=(2,0)\), while the two denominator factors generate
its left- and right-moving descendants.

Using the additive grading variables introduced in
Eq.~\eqref{eq:q_flow_coordinates}, the logarithmic partition function becomes
\begin{equation}
\log Z_{\log}(s,\bar s)
=
\sum_{k=1}^{\infty}
\frac{e^{-2ks}}
{k\left(1-e^{-ks}\right)
\left(1-e^{-k\bar s}\right)}.
\label{eq:log_partition_flow}
\end{equation}
Here
\begin{equation}
Z_{\log}(s,\bar s)
\equiv
Z_{\log}\left(e^{-s},e^{-\bar s}\right),
\label{eq:Zlog_pullback}
\end{equation}
so Eq.~\eqref{eq:log_partition_flow} is an exact
reparameterization of the established one-loop result in terms of
the additive conformal-grading variables. No spectral information is
changed by this rewriting.

In the domain
\begin{equation}
\Re s>0,
\qquad
\Re\bar s>0,
\label{eq:partition_convergence_domain}
\end{equation}
one has \(|q|<1\) and \(|\bar q|<1\), so the descendant factors admit
the geometric-series expansion
\begin{equation}
\frac{1}
{\left(1-e^{-ks}\right)
 \left(1-e^{-k\bar s}\right)}
=
\sum_{m,\bar m=0}^{\infty}
e^{-k(ms+\bar m\bar s)}.
\label{eq:descendant_geometric_expansion}
\end{equation}
Substitution into Eq.~\eqref{eq:log_partition_flow} gives
\begin{equation}
\log Z_{\log}(s,\bar s)
=
\sum_{k=1}^{\infty}
\frac{1}{k}
\sum_{m,\bar m=0}^{\infty}
e^{-k\left[(m+2)s+\bar m\bar s\right]}.
\label{eq:log_partition_descendant_expansion}
\end{equation}
This form makes the conformal grading explicit. The logarithmic
primary contributes the left-moving weight \(2s\), while its
descendants contribute \(ms+\bar m\bar s\). Equivalently, each term
carries the multiplicative grading
\begin{equation}
q^{\,k(m+2)}\bar q^{\,k\bar m}.
\label{eq:descendant_multiplicative_weight}
\end{equation}
Thus the primary and descendant contributions are organized by the
same additive parameters \((s,\bar s)\) that parameterize the
Virasoro zero-mode action. Equation~\eqref{eq:log_partition_flow}
therefore makes explicit the additive form of the conventional
multiplicative conformal grading without modifying the underlying
one-loop spectrum.

On the Euclidean conjugate locus introduced in
Eq.~\eqref{eq:s_u_theta_partition}, the convergence condition reduces to
\begin{equation}
u>0,
\label{eq:euclidean_partition_convergence}
\end{equation}
and Eq.~\eqref{eq:log_partition_flow} becomes
\begin{equation}
\log Z_{\log}(u,\theta)
=
\sum_{k=1}^{\infty}
\frac{e^{-2k(u+i\theta)}}
{k
\left[1-e^{-k(u+i\theta)}\right]
\left[1-e^{-k(u-i\theta)}\right]}.
\label{eq:log_partition_u_theta}
\end{equation}
The variable \(u\) controls the modulus of the conformal grading,
whereas \(\theta\) controls its phase. In particular,
\begin{equation}
q\bar q=e^{-2u},
\qquad
\frac{s+\bar s}{2}=u,
\end{equation}
so the coefficient governing the nilpotent mixing in the Virasoro
zero-mode action depends only on \(u\), while the partition function
retains \(\theta\)-dependence through its chiral semisimple grading
and descendant factors.

The bulk origin of Eq.~\eqref{eq:log_partition_q} remains the
one-loop determinant obtained from the gravitational fluctuation
problem \cite{Gaberdiel:2010xv}. The additive representation
Eq.~\eqref{eq:log_partition_flow} therefore provides an exact rewriting of the established one-loop result in terms of the additive parameters of the Virasoro zero-mode flow.

\subsection{Jordan evolution and graded state counting}
\label{subsec:jordan_evolution_state_counting}

The same Virasoro zero-mode operator \(U(s,\bar s)\) appears both in
the finite action on logarithmic states and in conformal state
counting, but the two uses probe different aspects of the
indecomposable representation.

Using the Jordan decomposition established in
Section~\ref{sec:virasoro_flow}, the restriction of the zero-mode
operator to the rank-two logarithmic pair is given by Eq. (\ref{eq:finite_complex_flow}). When acting on the logarithmic partner, the nilpotent term is retained explicitly as shown in Eq. (\ref{eq:flow_logarithmic}). This off-diagonal mixing is the representation-theoretic mechanism underlying the logarithmic dependence of the correlation functions derived in Section~\ref{sec:two_point_functions}.

Conformal state counting, by contrast, is organized by graded traces
of the zero-mode operator. Schematically,
\begin{equation}
Z(s,\bar s)
=
\operatorname{Tr}_{\mathcal H}
\left[
e^{-sL_0-\bar s\bar L_0}
\right],
\label{eq:statistical_trace_flow}
\end{equation}
where the usual vacuum-energy shifts have been suppressed since they
are not relevant to the present argument. Restricting the operator
to the illustrative finite rank-two Jordan block gives
\begin{equation}
\operatorname{Tr}_{\mathrm J}U(s,\bar s)
=
e^{-hs-\bar h\bar s}
\operatorname{Tr}_{\mathrm J}
\left[
\mathbf 1-\frac{1}{2}(s+\bar s)N
\right]
=
2e^{-hs-\bar h\bar s}
=
2q^h\bar q^{\,\bar h},
\label{eq:jordan_block_trace_joint}
\end{equation}
since \(\operatorname{Tr}_{\mathrm J}N=0\). Thus the ordinary trace
does not separately resolve the off-diagonal nilpotent matrix element
responsible for the Jordan mixing.

The factor \(2\) in Eq.~\eqref{eq:jordan_block_trace_joint} is simply
the dimension of the illustrative rank-two Jordan block,
\(\operatorname{Tr}_{\mathrm J}\mathbf 1=2\), and should not be
identified with a multiplicity in the logarithmic one-loop partition
function. In particular, for \((h,\bar h)=(2,0)\) the finite-block
trace is \(2q^2\), whereas the logarithmic sector entering
\(Z_{\log}\) has primary grading \(q^2\), as reviewed in
Section~\ref{subsec:log_partition_additive_coordinates}.

The state-space action and the graded trace therefore probe
complementary aspects of the logarithmic representation. Acting on
individual states, \(U(s,\bar s)\) explicitly retains the nilpotent
mixing, whereas the ordinary trace organizes conformal weights,
multiplicities, and descendant content without separately resolving
the off-diagonal Jordan matrix element.

\subsection{Relation to previous partition-function constructions}
\label{subsec:relation_previous_partition_constructions}

The logarithmic partition function employed here has previously been
studied from complementary combinatorial and geometric perspectives.
In Ref.~\cite{Mvondo-She:2018htn}, its plethystic expansion was used
to organize the multiparticle logarithmic sector and to expose
associated combinatorial structures, while
Ref.~\cite{Mvondo-She:2019vbx} developed related Schur-polynomial,
Hilbert-series, and symmetric-product interpretations. These
constructions concern structural information already encoded in the
same logarithmic partition function.

The point addressed here is different. The conventional
multiplicative grading variables \((q,\bar q)\) are related to the
additive parameters \((s,\bar s)\) through
\(q=e^{-s}\) and \(\bar q=e^{-\bar s}\), so that the established
partition-function grading is expressed in terms of the same
Virasoro zero-mode operator that governs the finite action on the
logarithmic pair. 

The present construction thus provides an exact
representation-theoretic reparameterization of the established
logarithmic partition function, relating its conformal grading to the
Virasoro zero-mode action. The previous combinatorial and geometric
interpretations remain unchanged and provide complementary
descriptions of the same logarithmic state-counting structure.

\section{Summary and outlook}
\label{sec:sum_and_out}

In this work, we have formulated the rank-two logarithmic sector of
critical Topologically Massive Gravity in terms of the Virasoro
zero-mode action
\begin{equation}
U(s,\bar s)
=
e^{-sL_0-\bar s\bar L_0}
=
e^{-hs-\bar h\bar s}
\left[
\mathbf 1-\frac{1}{2}(s+\bar s)N
\right],
\qquad
N^2=0.
\label{eq:summary_joint_action}
\end{equation}
For the critical-TMG logarithmic pair,
\((h,\bar h)=(2,0)\), with the equal nilpotent contribution
\(\frac{1}{2}N\) appearing in both \(L_0\) and \(\bar L_0\).
The finite action therefore separates naturally into a semisimple
conformal factor and a nilpotent mixing term governed by
\(\frac{1}{2}(s+\bar s)\).

Applied to the logarithmic pair, this zero-mode action reproduces the
characteristic coordinate dependence of rank-two LCFT two-point
functions. Under Euclidean conjugacy conditions,
\begin{subequations}
\begin{align}
\left\langle\psi^L\psi^L\right\rangle
&=0,
\\
\left\langle\psi^L\psi^{\log}\right\rangle
&=
b
\left(\frac{z_0}{z_{12}}\right)^{2h}
\left(\frac{\bar z_0}{\bar z_{12}}\right)^{2\bar h},
\\
\left\langle\psi^{\log}\psi^{\log}\right\rangle
&=
\left(\frac{z_0}{z_{12}}\right)^{2h}
\left(\frac{\bar z_0}{\bar z_{12}}\right)^{2\bar h}
\left[
d-b\log\left|
\frac{z_{12}}{z_0}
\right|^2
\right].
\label{eq:summary_log_correlators}
\end{align}
\end{subequations}
The semisimple action determines the conformal power law, whereas the
nilpotent Jordan mixing generates the logarithmic coordinate
dependence. These expressions reproduce the universal rank-two LCFT
structure, while the gravitational normalization and new anomaly
require the holographic bulk calculation
\cite{Skenderis:2009nt,Grumiller:2009mw}.

We have also related the additive parameters \((s,\bar s)\) to the
conventional multiplicative grading variables through
\(q=e^{-s}\) and \(\bar q=e^{-\bar s}\). This provides an exact
reparameterization of the established logarithmic contribution to
the one-loop partition function and places its conformal grading in
the same representation-theoretic framework as the Virasoro
zero-mode action. The finite state-space action and the ordinary
graded trace probe complementary aspects of the logarithmic
representation: the former explicitly resolves the off-diagonal
Jordan mixing, whereas the latter organizes conformal weights,
multiplicities, and descendant content without separately resolving
the nilpotent matrix element.

Several extensions are natural. For higher-rank logarithmic
representations, exponentiation of a nilpotent operator satisfying
\(N^r=0\) produces a finite polynomial of degree at most \(r-1\) in
the corresponding additive flow parameters. This suggests a
systematic extension of the present construction to higher-rank
logarithmic correlators, with critical generalized massive gravities
providing a natural setting in which such Jordan structures occur.

A further direction concerns the zero structure of finite-sector
partition functions obtained from the logarithmic state-counting
function. Since the multiplicative grading variables are related to
the additive zero-mode parameters by \(q=e^{-s}\) and
\(\bar q=e^{-\bar s}\), zeros in the grading plane may equivalently
be studied in the complexified additive parameter space. Their
organization and possible accumulation in an appropriate
large-sector limit could provide a route toward a Lee--Yang
description of the thermodynamic and analytic structure of the
logarithmic sector.

\paragraph{Acknowledgements} The author would like to thank Daniel Grumiller for his response to the author's query concerning the logarithmic contribution proportional to $\rho$ in the Fefferman-Graham expansion of the metric \cite{Grumiller:2008qz,Grumiller:2013at}. The author acknowledges financial support from the Department of Physics at the University of Pretoria.

\clearpage

\bibliographystyle{utphys}
\bibliography{sample}
\end{document}